# Orbital-collaborative Charge Density Wave in Monolayer VTe$_2$

Qiucen Wu[1], Zhongjie Wang[1], Yucheng Guo[1], Fang Yang[2,*], and Chunlei Gao[1,2,#]

[1] *State Key Laboratory of Surface Physics and Department of Physics, Fudan University, Songhu Rd. 2005, 200438 Shanghai, P.R. China*

[2] *Institute for Nanoelectronic Devices and Quantum Computing, Fudan University, Songhu Rd. 2005, 200438 Shanghai, P.R. China*

Corresponding authors: *fangyang@fudan.edu.cn, #clgao@fudan.edu.cn

**Abstract:**

Charge density waves in transition metal dichalcogenides have been intensively studied for their close correlation with Mott insulator, charge-transfer insulator, and superconductor. VTe$_2$ monolayer recently comes into sight because of its prominent electron correlations and the mysterious origin of CDW orders. As a metal of more than one type of charge density waves, it involves complicated electron-electron and electron-phonon interactions. Through a scanning tunneling microscopy study, we observed triple-Q $4 \times 4$ and single-Q $4 \times 1$ modulations with significant charge and orbital separation. The triple-Q $4 \times 4$ order arises strongly from the p-d hybridized states, resulting in a charge distribution in agreement with the V-atom clustering model. Associated with a lower Fermi level, the local single-Q $4 \times 1$ electronic pattern is generated with the p-d hybridized states remaining $4 \times 4$ ordered. In the spectroscopic study, orbital- and atomic- selective charge-density-wave gaps with the size up to $\sim 400\ meV$ were resolved on the atomic scale.

Charge density waves (CDWs) in transition metal dichalcogenides (TMDs), $MX_2$ (M=Ti, V, Nb, &Ta, X=S, Se, &Te), have provoked fundamental interests in electron-phonon coupling and electron correlations [1-4]. Various states of matters, such as insulator, superconductor, and normal metal, are closely related to the CDW orders in the phase diagram [5-12]. Modeling a specific CDW material in theory is highly challenging owing to numerous factors, such as hybridizations, Coulomb interactions, band structures, and so on. The periods of many CDW phases associated with structural modulations in TMDs can be explained by the Fermi surface nesting of the d electron bands [13]. As an exception, the $d^1$ configuration favors an atom clustering, which can be understood by the chemical bonding of d orbitals [13]. A typical example is the $\sqrt{13} \times \sqrt{13}$ CDW ensembled by star-of-David metal atom clusters, which has been deeply studied in $TaS_2$ and $TaSe_2$ [7,11,14,15]. Recently, the $VTe_2$ monolayer (ML) was found to be of the $d^1$ configuration with $4 \times 4$ CDW [16]. In experiment, an angle-resolved photoemission spectroscopy (ARPES) study revealed that the CDW order opens an anisotropic gap near the Fermi level [17]. The microscopic images of CDW modulations obtained with scanning tunneling microscope (STM) show different phase shifts relative to the lattice implying an inter-unit-cell distribution of orbitals [16,18-20]. These experimental findings suggest a nonuniform distribution of the CDW orders in real space. In theory, the phonon calculation came out with the instability of quadruple-unit-cell period [17,18], while the density functional theory (DFT) calculation obtained a preference of the $3 \times 1$ structural modulation similar to the bulk [16,18]. Complex CDW phenomena generally involve competing interactions. As the $VTe_2$ ML has not only d and but also p bands near the Fermi level [16-18,21], the structural modulation may result from the Jahn Teller effects driven by the p orbitals [13]. In addition, the weak Coulomb screening in 2D materials would give rise to large binding energy of excitons and a nonlocal interaction of charge [22]. Thus, to fully understand the CDW mechanism of $VTe_2$ ML, a thorough study combining orbital, charge and CDW order emergent from the electronic instability is highly demanded.

In this letter, we investigated the $VTe_2$ ML grown on the highly oriented pyrolytic graphite (HOPG) substrate, where the $4 \times 4$ CDW has been discovered in previous reports [16-20]. Our results demonstrate that the universal $4 \times 4$ modulations are bonded to the p-d hybridized states. A spatial charge separation of p and d states is resolved within the $4 \times 4$ unit cell on the atomic scale. These findings are further verified by comparing the electronic structure of the $4 \times 4$ CDW with an additional $4 \times 1$ modulation. We suggest that the p-d hybridization and the nonlocal Coulomb interaction should be particularly considered in such a big CDW unit cell.

The experiments were carried out in a combined molecular beam epitaxy and low-temperature STM system with a base pressure of $1 \times 10^{-10}$ $mbar$. The growth of $1T-VTe_2$ was performed at a substrate temperature of 315 °$C$ with a growth rate of 30 mins/ML. Through STM measurement at $5\ K$, we found that the low-temperature CDW phases of the ML 1T-VTe$_2$ are sensitive to the growth temperature of the substrate. The structural modulations in previous reports [19,23] may depend on growth conditions as well. The samples that we will discuss in this letter have only areas with $4 \times 4$ and $4 \times 1$ modulations.

Fig. 1(a) shows the topographic image of ~0.8 ML VTe$_2$ thin films on the HOPG substrate. The first layer forms continuous islands covering most of the substrate and the second layer forms small triangular islands. As is well known, the first layer undergoes a CDW phase transition at ~190 K from a pristine 1T phase to a $4 \times 4$ CDW phase [17,20]. The zoomed-in images of the first layer exhibit voltage dependent $4 \times 4$ and $4 \times 1$ modulations as shown in Fig. 1(b) and 1(c). Surprisingly, while the period in the $4 \times 4$ area remains the same at different voltages, the $4 \times 1$ area turns to be more $4 \times 4$-like at a negative voltage in Fig. 1(c). The atomically sharp boundary between $4 \times 4$ and $4 \times 1$ areas evidences the metastability of the two phases.

Fig. 1(d) and 1(e) show the atom resolved images of $4 \times 4$ and $4 \times 1$ areas, respectively. In the $4 \times 4$ area, the upper Te layer is imaged with a $C_{3v}$ symmetry at the surface. In the $4 \times 1$ area, the upper Te layer forms stripes with different brightness, where a $2 \times 1$ period stands out. Consequentially, the Fourier transform [Fig.1 (g)] of Fig. 1(e) displays a higher intensity of the $2 \times 1$ period than the $4 \times 1$ one. Similarly, the high order spots of the $4 \times 4$ periods in Fig. 1(f) show comparable intensities to the first order. In spite of the distinct topography in the two areas, their averaged tunneling spectra are rather similar as shown in Fig. 1(h). But, the negative-voltage spectrum of the $4 \times 1$ area has an overall shift of $\sim 30\ mV$ relative to the $4 \times 4$ one, while the positive-voltage spectra show a peak (marked by a black arrow) at $\sim 0.3\ V$ only in the $4 \times 1$ area.

The $dI/dU$ values as functions of energy and position are obtained by mapping the tunneling spectra in both $4 \times 1$ and $4 \times 4$ areas (Fig. S3). The $dI/dU$ cuts at selected energy are displayed in Fig. 2(e-p). To learn general information about the electronic structure, the inter-energy cross-correlation $C(E_1, E_2)$ in both areas was calculated to characterize the similarities between the $dI/dU$ maps at a different energy, which is similar to the method applied in Y. Chen et al.'s work [24]. The formula to calculate cross-correlation is:

$$C(E_1, E_2) = \frac{\sum_{x,y}\{[\rho(E_1,x,y) - \overline{\rho(E_1)}][\rho(E_2,x,y) - \overline{\rho(E_2)}]\}}{\sqrt{\sum_{x,y}[\rho(E_1,x,y) - \overline{\rho(E_1)}]^2 \sum_{x,y}[\rho(E_2,x,y) - \overline{\rho(E_2)}]^2}}$$

where $\rho(E_{1,2}, x, y)$ is proportional to the $dI/dU$ value measured at the bias voltage $U_{1,2} = E_{1,2}/e$ ($e$ represents the electron charge) and the spatial coordinates $(x, y)$. $C(E_1, E_2) = 1\ (-1)$ indicates the exactly same (opposite) pattern at energy $E_1$ and $E_2$. $C(E_1, E_2)$ in $4 \times 4$ and $4 \times 1$ areas are plotted in Fig. 2(a) and 2(c), respectively. The correlation between the occupied states shows checkboard patterns in both areas with a higher Fermi level in the $4 \times 4$ area. Hence, we could divide the $dI/dU$ patterns into two classes, e.g., Fig. 2(m) and 2(o) for one class and Fig. 2(n) and 2(p) for the other. Comparing Fig. 2(a) and 2(c) to the band structure extracted from ref. [21] [Fig. 2(b)], the voltage evolution of the cross-correlation keeps pace with the switching of dominant orbitals between d (green frames) and p (blue frames). In other words, the two classes of $dI/dU$ patterns represent the spatially anticorrelated p and d dominating states. The coherent density-of-states (DOS) distribution of p and d orbitals suggests an orbital hybridization in the orbital overlapping energy of $-110 \sim -710\ meV$, which would open a gap at the band crossing near $-500\ meV$ in the $\Gamma - M$ direction [Fig. 2(b)] as consistently observed by ARPES [16-18,21].

In the $4 \times 4$ area, a V-atom clustering model is provisionally considered with the center at the "A" site and the other sites ("B", "C", and "D") moving towards the nearest "A" site in the $4 \times 4$ lattice displayed in Fig. 2(d). At the meantime, five Te sites (labeled as "1", "2", "3", "4", and "5") are defined according to their positions with respect to the V atoms. Compared to the STM images, the "5" locates at the center of six brightest "2" atoms in the $4 \times 4$ unit cells in Fig. 1(d) and the "A" site is adapted to the local $dI/dU$ maxima at $-170\ mV$ in Fig. 2(m) as well as the highest protrusions in Fig. 1(c). As a simple verification of this V-clustering model regarding the lattice symmetry, the d states of the equivalent V atoms marked in Fig. 2(d) should contribute to the same $dI/dU$ values in the $4 \times 4$ area, e.g., the bright (yellow) atomic sites at positive voltages: "A" and "C" sites in Fig. 2(e-f), "B" and "D" sites in Fig. 2(g-h), "D" sites in Fig. 2(i), "C" sites in Fig. 2(j), and "C" sites in Fig. 2(k). In the $4 \times 1$ area, the nonoverlapping p and d orbitals at $> -110\ meV$ or $< -710\ meV$ are ordered in stripes except for $4 \times 4$ modulations [Fig. 2(j)] in a narrow energy range $190 \sim 230\ meV$ (Fig. S3) (will be discussed later). However, the in-phase $4 \times 4$ modulations are found to cover both areas in the p-d overlapping energy (Fig. S3). As expected, the $dI/dU$ maps in the energy range corresponding to the blue frames around $-400\ meV$ in Fig. 2(a) and 2(c) show the opposite contrast to the green frames. In brief, it is surprisingly found that the $4 \times 4$ modulation is bonded to the orbital hybridized state regardless of the spatial phase separation.

As consequences of the V clustering, the largest charge density at "A" sites would lead to the highest Fermi level and the largest DOS of d orbital. To trace the spatial variation of Fermi level and DOS, the spectra are extracted along the "L" shape [Fig. 2(m)] passing the V sites as "A-B-D-B-A-C-D-C-A" in both $4 \times 4$ and $4 \times 1$ areas. The results are given in Fig. 2(q) and 2(r). It is noted that the voltage shift of the peaks at $\sim -0.2\ V$ agrees with the Fermi level lowering in the sequence of "A", "B", "C"≈"D" in both $4 \times 4$ and $4 \times 1$ areas. The relatively larger voltage shift of the peaks at $\sim -0.6\,V$ is further influenced by the spatial and energy distribution of p and d orbitals as illustrated in Supplementary Materials S2. Meanwhile, both of the peaks maximize at the "A" site as expected for the highest DOS.

In Fig. 2(q) and 2(r), two groups of peaks correlated in space and energy (marked by black and red arrows) signify the local pairs of valence bands and conduction bands. First, the black arrows mark a valence band ($\sim -150\,meV$) and a weak conduction band ($\sim 300\,meV$) in the $4\times 4$ area. With a noticeable lower Fermi level in the $4\times 1$ area, the conduction band develops into a remarkable peak, while the valence band gets suppressed. Such a DOS transfer over the Fermi level implies the band gap originated from the Coulomb interaction. In the aforementioned energy range of $190\sim 230\,meV$, the bottom part of the conduction band shows a similar $4\times 4$ modulation to the valence band reflecting a degeneracy lift of the p-d hybridized state. In a higher energy range of $250\sim 390\,meV$, the conduction band turns out to be $4\times 1$ stripes as a typical feature for the nonhybridized state. Second, the red arrows mark a valence band ($\sim -50\,meV$) and two conduction bands ($\sim 50\,meV$ and $\sim 130\,meV$) in the $4\times 4$ area. In the $4\times 1$ area [Fig. 2(q)], only the peaks at $\sim \pm 50\,meV$ could be resolved with a weak contrast. In the transition from the $4\times 4$ to the $4\times 1$ area, the suppression of CDW features marked by red arrows seems to have the opposite fate to the enhancement of those marked by black arrows. In Fig. S1(a), the spatial evolution of these two CDW gaps near the boundary is visualized with higher energy and spatial resolution. These evidences bridge the CDW orders with the Fermi level.

The energy narrowness of the valence and the conduction bands describes the strongly localized gaps or pseudogaps. Although it is beyond the lateral resolution of STM to identify whether a gap is fully developed on the atomic scale, a maximally 70% spatial corrugation of the occupied DOS (Fig. S3) illustrates a definitely strong charge localization. An anisotropic CDW gap has been reported by the previous ARPES study with a $50\,meV$ maximal Fermi energy [17]. In agreement with the gap anisotropy in the momentum space, our results reveal that the CDW gaps are strongly orbital and atom-site dependent with the size up to $\sim 400\,meV$ (peak-to-peak).

The above discussion clearly demonstrates that the p-d hybridized states generate a triple-Q $4\times 4$ modulation, while the single-Q $4\times 1$ pattern appears locally as a stripe ordering of all p and d bands imposed on the $4\times 4$ order. Owing to the complex electronic structure near the Fermi level, it is difficult to pin down the origin of the single-Q order. Here, we will put forward several relevant aspects to have an open discussion on the emergence of $4\times 1$ modulations. First, the occurrence of the single-Q order is in company with a reduction of charge density, i.e., the occupied electron states. Provided the preference of quadruple-unit-cell susceptibility of the VTe$_2$ ML as predicted by phonon calculations [17,18] and the p-d charge separation that we have observed, the p-d staggered $4\times 4$ modulations is the optimal solution for the CDW order subjected to the Coulomb energy. With a reduced Coulomb interaction, modulations of other periods may come out, e.g., the $4 \times 1$ modulation. Second, the V-clusters would have ground states of degenerate chemical bonds similar to the star-of-David clusters in TaS$_2$ [11,15,25]. A lower Fermi level means the smaller electron filling numbers in the degenerate bonds. To some extent, the electrons would fill some of bonds in prior to the others, which leads to a single-Q ordering. The similar origin has been used to explain many single-Q CDW orders in TMDs [13]. Third, the VTe$_2$ layers show an intension to form stripe orders, since a quite complete collection of periods has been discovered, such as $2 \times 1$ for few layers [18,26], $3 \times 1$ for bulk [27,28], and $4 \times 1$, $5 \times 1$ for ML [19]. Forth, the single-Q order seems to be sensitive to the charge doping in TMD materials. For example, The triple-Q to single-Q transition was observed in TaS$_2$ with Hf substitution [29] and TiSe$_2$ with Cu intercalation [30]. However, the local pinning and the band doping effects of the defects cannot be completely disentangled in the VTe$_2$ ML.

In conclusion, we investigated VTe$_2$ ML on the HOPG substrate with $4 \times 4$ CDW and local $4 \times 1$ patterning. Our results reveal the distinct electronic features contributed by $4 \times 4$ and $4 \times 1$ orders. The interorbital interaction contributes to an anticorrelated charge distribution of p and d orbitals, which supports the formation of the $4 \times 4$ CDW ordering. On the other hand, the local variation of CDW gaps, along with the change of the Fermi level from the $4 \times 4$ to the $4 \times 1$ area, would lead to further discussions about the interplay of Hund's coupling and crystal field splitting [31,32].

We thank Dr. Wei Ruan for insightful discussion. F.Y. is sponsored by Shanghai Pujiang Program No. 19PJ1401000. C. G. acknowledges the funding by the National Natural Science Foundation of China (Grant Nos. 11427902, 11674063), National Key Research and Development Program of China (Grant No. 2016YFA0300904).

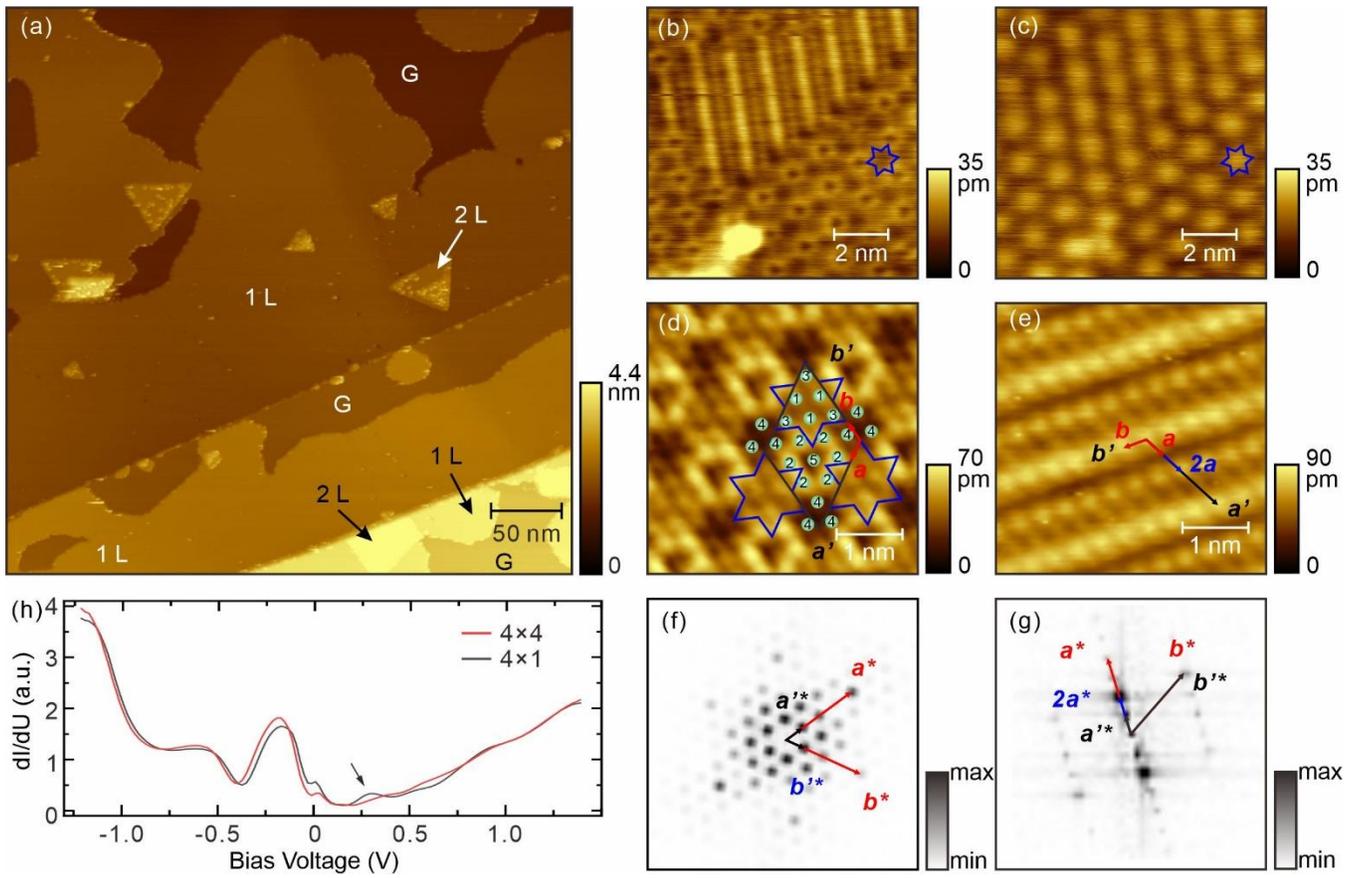

Fig.1 (a) STM image of first and second layer VTe$_2$ islands ($U = 1.3\,V, I = 100\,pA$). (b, c) STM image of the same area on ML VTe$_2$ [(b) $U = 0.5\,V, I = 100\,pA$ and (c) $U = -0.3\,V, I = -80\,pA$]. (d) [(e)] High-resolution STM image of $4 \times 4$ ($4 \times 1$) areas with the tip-sample distance stabilized at $U = 0.5\,V, I = 100\,pA$ and (f) [(g)] the Fourier transform images. In (d) and (e), $\boldsymbol{a}$, $\boldsymbol{b}$ represent the primitive vectors and $\boldsymbol{a'}, \boldsymbol{b'}$ represent the lattice vectors of $4 \times 4$ and $4 \times 1$ modulations. In (f) and (g), their reciprocal lattice vectors are denoted as $\boldsymbol{a}^*$, $\boldsymbol{b}^*$, $\boldsymbol{a'}^*$, and $\boldsymbol{b'}^*$ accordingly. (h) Spatially averaged tunneling spectra of $4 \times 4$ and $4 \times 1$ areas.

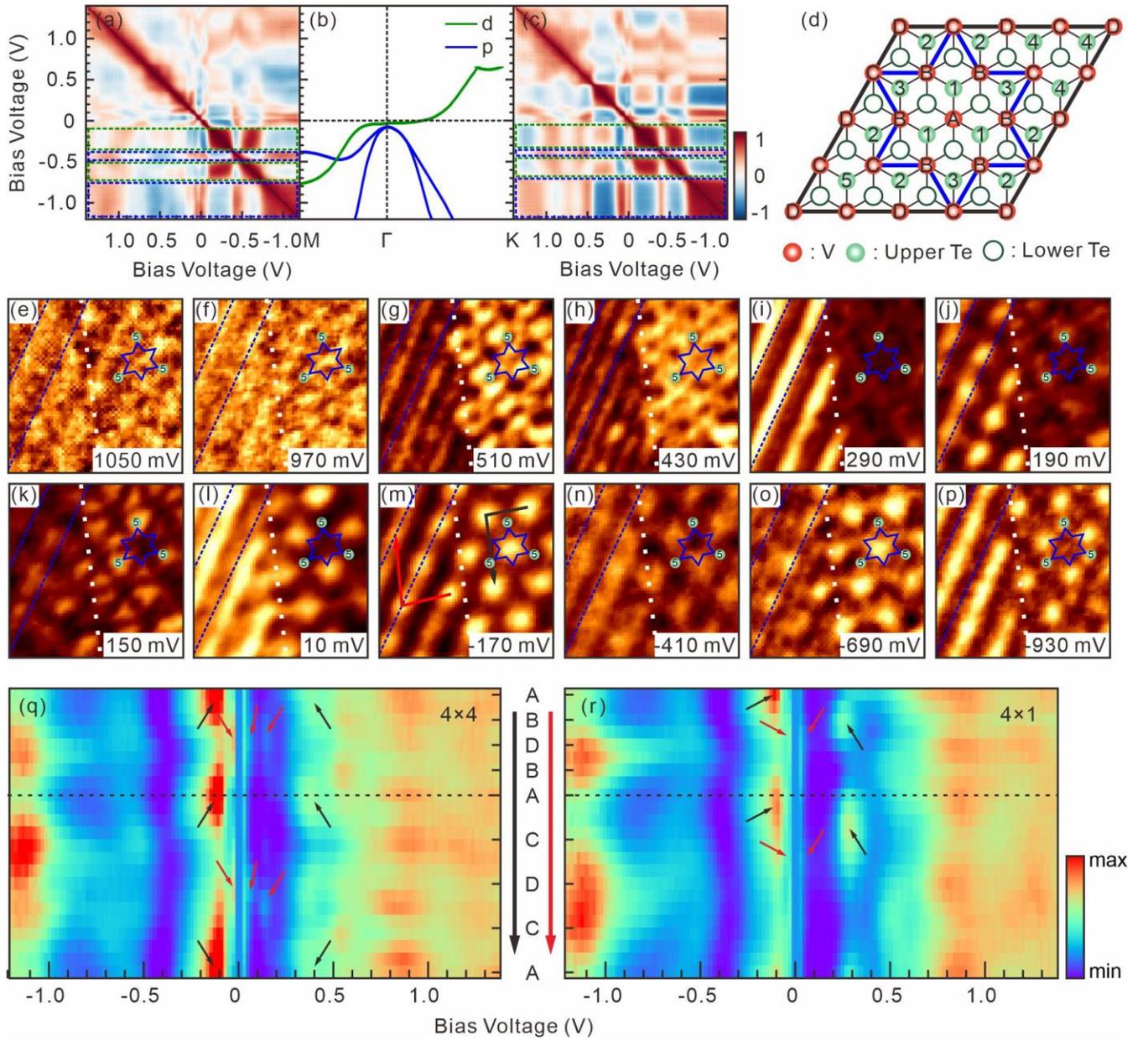

Fig.2 (a–c) The inter-energy cross-correlation in (a) $4 \times 4$ and (c) $4 \times 1$ areas and (b) the band structure extracted from Ref. [21] with the d bands in green and the p bands in blue. (d) The V-clustering model portrayed in the $4 \times 4$ atomic lattice of ML 1T-VTe$_2$. (e-p) $dI/dU$ maps of the same area with $4 \times 4$ and $4 \times 1$ modulations at labeled bias voltages with the tip-sample distance stabilized at $U = 1.4\,V$, $I = 1.0\,nA$ and bias modulation $\Delta U_{rms} = 20\,mV$ and $f = 983\,Hz$. The white dashed line indicates the boundary between the $4 \times 1$ (left) and $4 \times 4$ (right) areas. The star-of-David and the Te-site "5" in (d) are pasted in (e-p) at the corresponding positions. (q)[(r)] The $(dI/dU)/(I/U)$ spectra extracted from $4 \times 4$ ($4 \times 1$) areas at positions along the black (red) arrowed "L" shaped traces in (m). The turning points of the traces are marked by black dotted lines in (q) and (r). The positions of the atom sites are indicated beside (q) and (r).

# Supplementary Material for "Orbital-collaborative Charge Density Wave in Monolayer VTe$_2$"

## S1. Spatial evolution of tunneling spectra from the $4 \times 4$ to the $4 \times 1$ area.

Fig. S1(a) shows the topographic image of both $4 \times 1$ and $4 \times 4$ areas. The boundary of these two areas is marked as a dashed line. The tunneling spectra were recorded along the blue arrowed line in Fig. S1(a) with a high spatial resolution. To enhance the contrast, the $dI/dU$ values are plotted as functions of energy and displacement [the right panel of Fig. S1(b)] after subtracting the minimum of each energy [the left panel of Fig. S1(b)]. The occupied states show an obvious upshifting from the $4 \times 4$ to the $4 \times 1$ area representing the Fermi level shift. The red and black sets of the CDW gaps are marked by the colored arrows in Fig. S1(b) similar to Fig. 2(q) and 2(r). Note that the local maxima in Fig. S1(a) are not on "A" sites at a positive bias voltage. Nevertheless, the peaks for valence bands and conduction bands show a nice correlation in real space. With a higher energy resolution compared to Fig. 2, more fine structures in spectra are visualized. For example, the valence bands marked by red arrows split into two near the boundary.

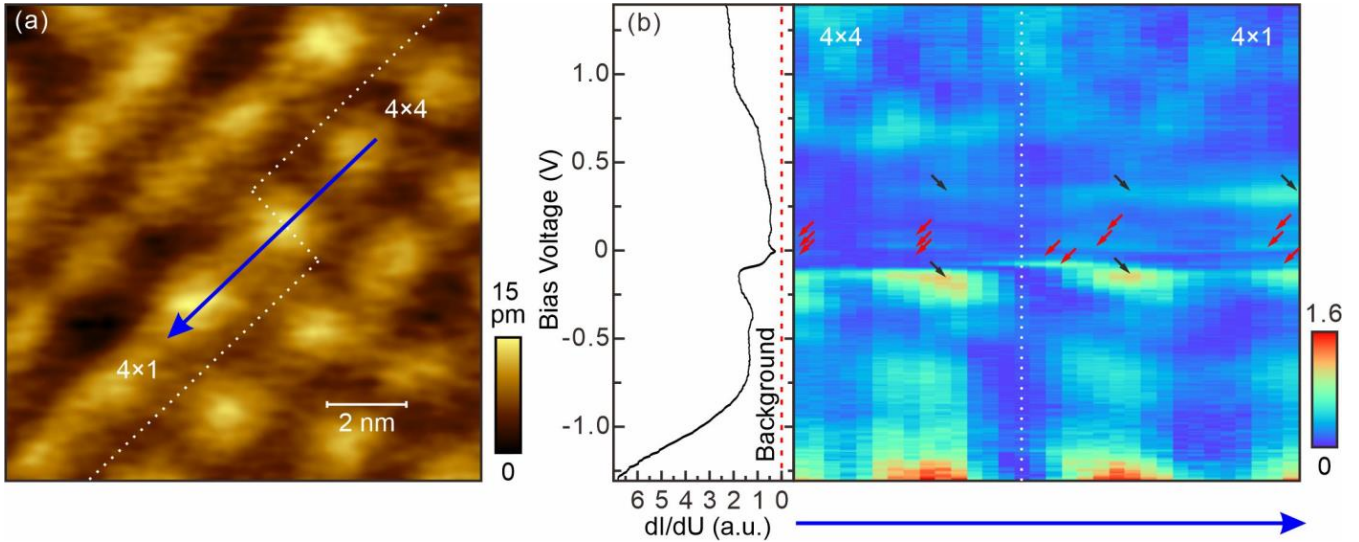

Figure S1. (a) STM image including $4 \times 4$ and $4 \times 1$ areas and their boundary (white dashed line) ($U = 1.4\ V, I = 500\ pA$). The blue arrow exhibits where STS line spectra were acquired. (b) STS spectra acquired along the blue arrow in (a) with bias modulation $\Delta U_{rms} = 7\ mV$ and $f = 983\ Hz$.

## S2. Tunneling spectra at specific positions in the CDW unit cell

Fig. S2(a) shows the topographic image of the area where all data were measured in Fig. 2. With a lower positive voltage, the topography of the $4 \times 1$ area is dominated by the stripe-like features at $\sim 0.3\ V$. The position of the "A" sites is located at the $dI/dU$ maxima in the $4 \times 4$ modulation at $-170\ mV$ in Fig. 2(m). In the $4 \times 4$ area, the averaged tunneling spectra of the atoms with the same atomic numbers in Fig. 2(d) are plotted in Fig. S2(b). "X" denotes the position of the largest charge density of the pure p orbital indicated in Fig. 2(p), which is in the center of three "A" sites other than "5" [Fig. 2(d)]. In the $4 \times 1$ area, the averaged tunneling spectra of the atoms along the same stripes are plotted in Fig. S2(c). The stripe "α" is defined as the brightest stripe at $290\ mV$ in the $4 \times 1$ area of Fig. 2(i). The other 3 stripes are spaced by one lattice vector labeled as "β", "γ", and "δ". The averaged spectra of "A" sites and "X" sites in the $4 \times 1$ area are plotted as well in Fig. S3(c) for comparison.

In the $4 \times 4$ area [Fig. S2(b)], the intensities of the peaks near $-0.2\ V$ decrease and their positions shift to the right in the sequence of "A" to "B" to "C" and "D" indicating the charge distribution due to the V-atom clustering. As expected for the charge maximum of the p orbitals, the "X" site shows the highest intensity at voltage below $-0.8\ V$. In the range from $-0.5$ to $-0.7\ V$, the peaks of "A" and "B" sites are located at lower voltage than those of "C", "D", and "X" sites. The former represents the energy of the flat d-band and the latter represents that of the flat p-band near the $M$ point in Fig. 2(b), which is consistent with the charge separation of p and d orbitals.

In the $4 \times 1$ area [Fig. S2(c)], the "A" site makes the highest peak near $-0.2\ V$, because both the $4 \times 4$ and the $4 \times 1$ ordering give rise to its high charge density. The "α" stripe contributes the pronounced $dI/dU$ peak of d orbital at $\sim 0.3\ V$ as in Fig. 2(i), and the "γ" stripe dominates the voltage below $-0.8\ V$ for its p orbital nature as in Fig. 2(p). Other features basically agree with the smooth transition of the aforementioned features and superimposed features from the $4 \times 4$ order.

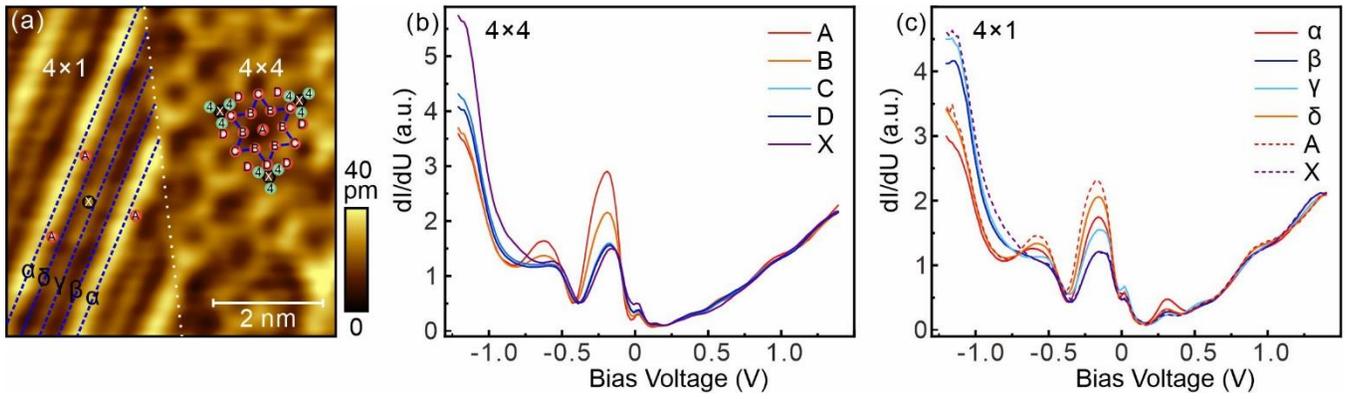

Figure S2. (a) STM image of $4 \times 4$ and $4 \times 1$ areas on ML 1T-VTe$_2$ ($U = 0.5\ V$, $I = 100\ pA$). In the $4 \times 4$ area, the atom sites are numbered in the same way as Fig. 2(d-p). The "X" site is the lower Te site surrounded by three "4" upper Te sites. In the $4 \times 1$ area, four stripes are labeled as "α", "β", "γ", and "δ". The "α" stripe is located at the brightest stripes in Fig. 2(i). The "A" site is marked according to Fig. 2(m), and "X" is positioned at the center of the three "A" sites. (b) Tunneling spectra averaged over all equivalent atom sites "A", "B", "C", "D", and "X" in the $4 \times 4$ region. (c) Tunneling spectra averaged over all equivalent atom sites on stripes ("α", "β", "γ", and "δ") and atom sites ("A" and "X") in the $4 \times 1$ area.

## S3. Energy dependences of $dI/dU$ patterns and corrugation

In this section, we present more data from the measurements presented in Fig. 2. The first (top left corner) figure is the topographic image, i.e., the tip-sample distance map. The corrugation at this setpoint condition ($U = 1.4\ V, I = 1.0\ nA$) comes out with a standard deviation of $\delta = 2.8\ pm$ and a total range of $19.0\ pm$, which is close to the constant height condition compared to the topography with other bias voltages. The white dashed line in the topographic images is used to divide $4 \times 1$ and $4 \times 4$ areas. The blue dashed line and the star-of-David sign are defined as the same in Fig. 2. In the $dI/dU$ maps, the corrugations of the maps are defined as:

$$\sigma = \frac{dI/dU_{max} - dI/dU_{min}}{dI/dU_{max} + dI/dU_{min}}$$

It is worth noting that the corrugation of $dI/dU$ at high positive voltages ($> 0.81\ V$) is smaller than 0.2, while it is generally above 0.4 at other energy. Two flat d bands stay at high positive energy, of which the symmetry is different from the one crossing the Fermi level. So, the high-energy d bands are not sensitive to the structure distortions caused by the electronic instability near the Fermi level. The partially filled p and d bands show up to 0.74 corrugations for a strong charge localization.

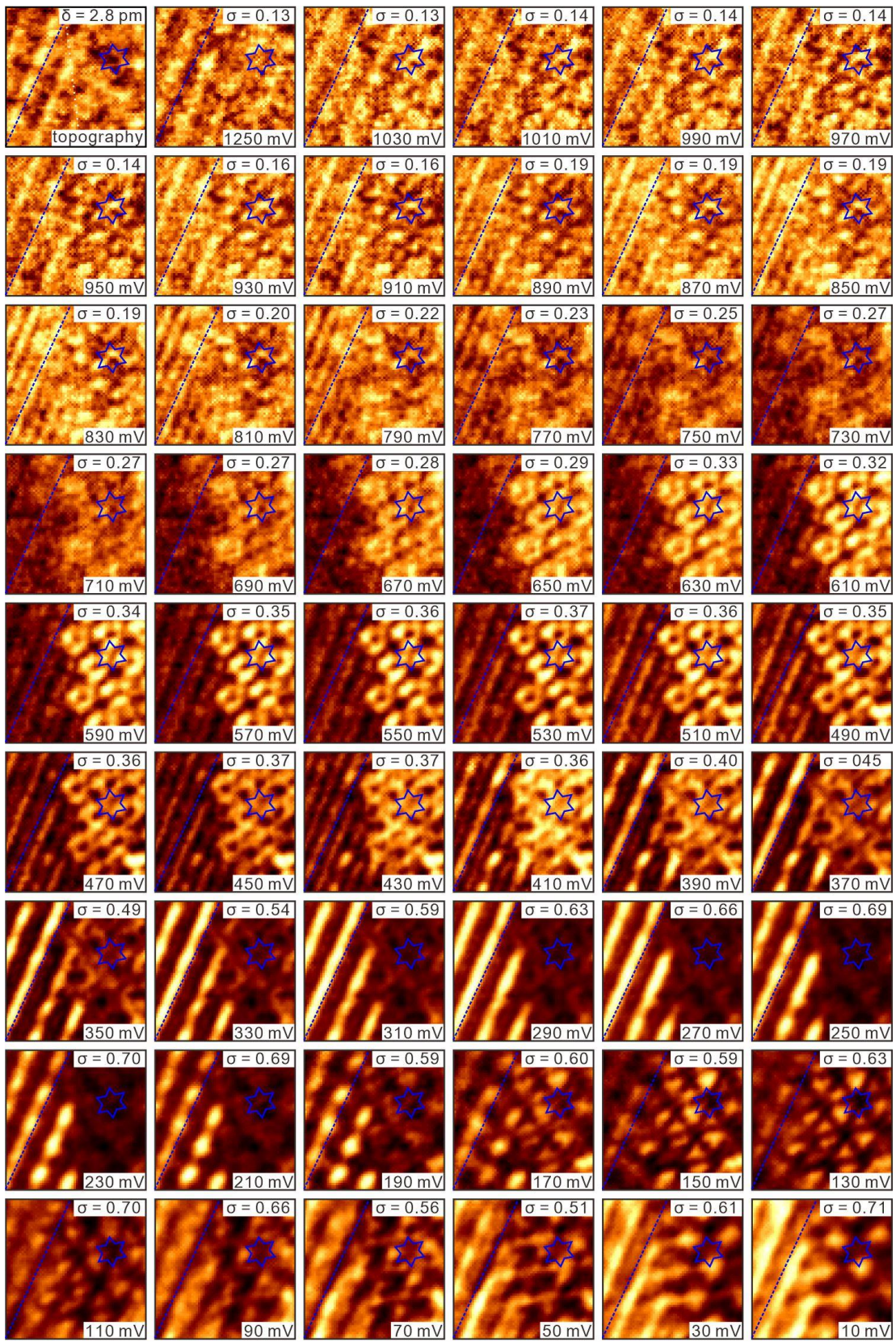

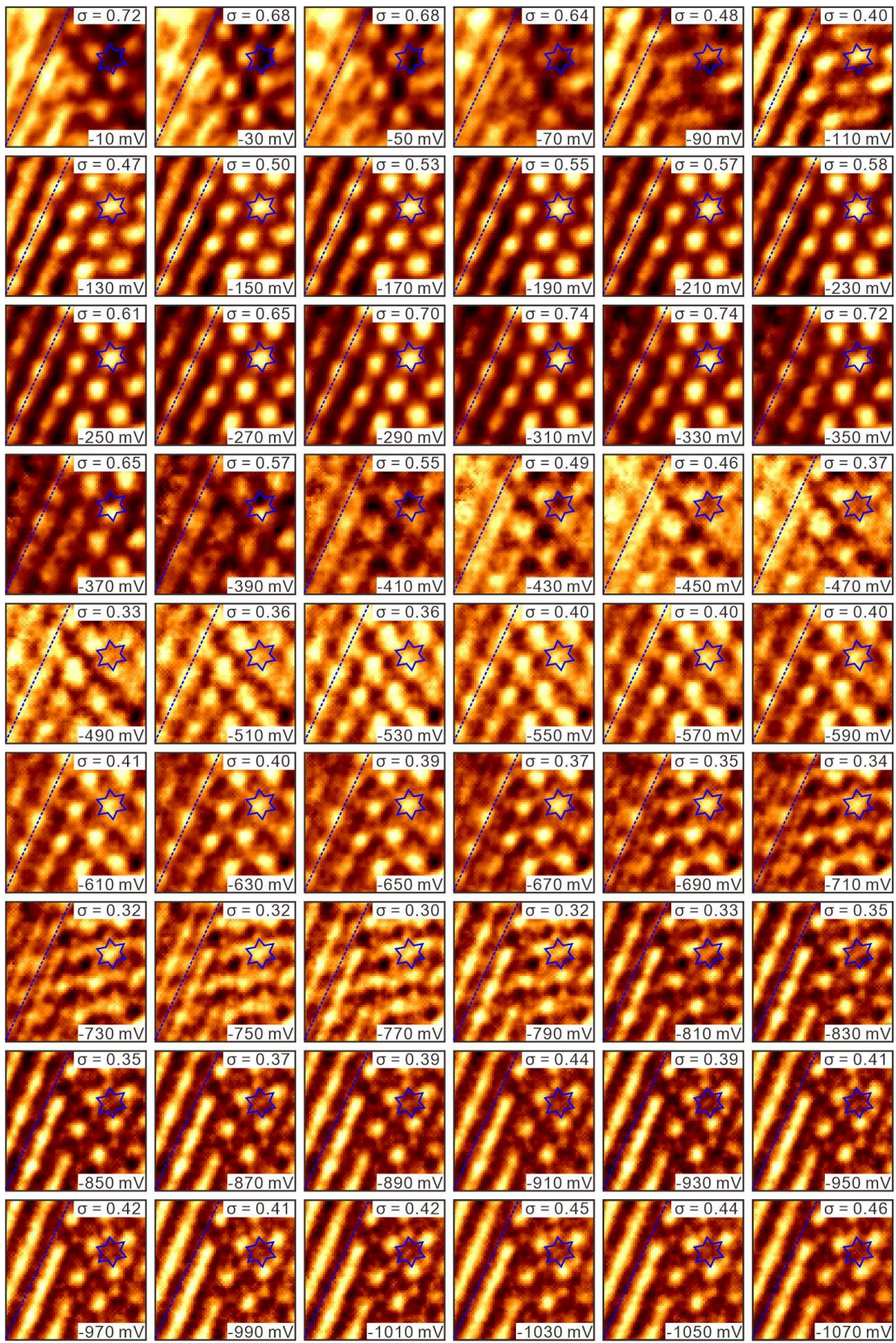

Figure S3. Topographic image and $dI/dU$ maps of the data in Fig. 2. The voltage and corrugation are marked on the images.